\documentstyle[preprint,prd,aps]{revtex}
\begin{document}
\draft
\preprint{
\vbox{
\halign{&##\hfil\cr
	& AS-ITP-98-09 \cr
        & hep-ph/9806505 \cr
	& Revised in Oct. 1998 \cr}}
}
\title{Bilinear R-parity Violation and $\tau^{\mp} \tilde{\kappa}^{\pm}$ 
mixing production in $e^{+} e^{-}$ colliders}
\author{Chao-Hsi CHANG$^{1,2}$, Tai-Fu FENG$^2$ and Lian-You SHANG$^2$}
\address{$^1$CCAST(World Laboratory), P.O.Box 8730,
Beijing 100080, P. R. China}
\address{$^2$Institute of Theoretical Physics, Academia Sinica, P.O.Box 2735,
Beijing 100080, P. R. China}
\address{Email: fengtf@itp.ac.cn}
\maketitle
\begin{center}
\begin{abstract}
A R-parity breaking SUSY model characterized by an effective
bilinear violation with only $\tau$-lepton number
breaking in the superpotential is outlined. The 
CP-odd Higgs boson masses and those of charged Higgs bosons are
discussed. In the model, several interesting
mass mixings else such as the mixing between $\tau$ lepton
and charginos etc in the model are discussed too. 
Being one of example, we have computed
the mixing production $e^{+} e^{-} \rightarrow \tau^{\mp} 
\tilde{\kappa}^{\pm}_{i} (i=1,2)$ in $e^{+} e^{-}$ colliders, 
where $\tau^{\mp}, \tilde{\kappa}^{\pm}_{i}$($i=1,2$)
denote the physical $\tau$ lepton and charginos.

\end{abstract}
\end{center}

\section{Introduction}
The minimal supersymmetric extension of the Standard 
Model(MSSM)\cite{s1} is believed one of the most attractive 
candidates beyond the Standard Model(SM) now. 
In the usual MSSM an additional quantum
number, the so-called R parity of a particle: $R=(-1)^{2S+3B+L}$\cite{s2},
is assumed to be conserved, where, besides the spin
quantum number S, L is the lepton number and B is the baryon number. 
In such a case with R-parity conserved,
all supersymmetric particles must be pair-produced, while the 
lightest of super-partners must be stable. Whether or not
with a conserved R-parity, the supersymmetric realization
is an open dynamical question, sensitive to physics at a 
more fundamental scale\cite{s3}.
Whereas if relaxing the R-parity conservation and
the relaxing will not conflict
with all the observations such as the proton decay and the other
rare decays for quarks, leptons and weak bosons etc., we may 
have new insight to see the long standing problems of particle 
physics, such as the neutrinos masses problem etc and can make
the supersymmetric realization to occur at a comparative lower
energy scale. Remarkably, for instance
the neutrino can acquire the tree level supersymmetric masses via 
the mixing with the neutralinos at the weak scale in the R-parity
violation framework\cite{s4,s5,s6,s7,s8}. 
This mechanism does
not involve in the physics at the large energy scale $M_{int} \sim O(10^{12}
GeV)$. It is, in contrast to the see-saw mechanism, relate the neutrino mass
to the weak-scale physics that is more accessible for experimental 
observations\cite{s9}.

The R-parity can be broken explicitly\cite{s10} or 
spontaneously\cite{s11}, that depends on the superpotential
and the soft SUSY breaking pattern precisely of the model. 
The first option allows one to 
establish very general phenomenological consequences 
of R-parity violation while the second one, R-parity is 
kept at the Lagrangian level as a
fundamental symmetry, but it is broken by the 
vacuum i.e. the ground state of the world. 
For the second, there are quite a lot of possible
virtues being added, such as a possibility of having a dynamical 
origin for the breaking of R-parity through radiative 
corrections if SUSY has been broken already, that is very 
similar to certain models for the electroweak symmetry
breaking\cite{s12} etc.

In this paper we focus on the truncated version of such a model, namely
in which the violation of R-parity is effectively introduced by a
bilinear superpotential term $\epsilon^{I} \varepsilon_{ij} \hat{L}^{I}_{i}
\hat{H}^{2}_{j}$ with proper soft SUSY breaking pattern that the R-parity
violation is not only originated to the precise R-parity breaking term in the 
superpotential but also to the vacuum. Here $\hat{L}^{I}(I=1,2,3)$ denote 
three generations of the 
leptonic SU(2) supersymmetric fields, thus the 
term also breaks the leptonic numbers. Whereas we will 
assume that only one generation of the lepton number, i.e.
the $\tau$ lepton number, is broken for simplicity. To deduct 
free parameters in the model so `artificially' by the assumption
here is because we may argue and believe the third generation is special
based on the fact that the third generation is very heavy, especially,
the top-quark mass so heavy $m_t\simeq 175 GeV$ close to
the electroweak broken scale already. In addition, we think 
the general feature of the model in phenomenology can still be kept, 
even the leptonic numbers of the other two generations are broken 
occasionally in the same way. In this effective truncated model,
the all superfield contents are exactly the same as those of the
MSSM but the R-parity violation is broken and
realized by the bilinear R-parity violation in the
superpotential. Generally the superpotential and
the relevant soft breaking terms of the model may also
lead to two scenarios: the vacuum expectation values(VEVs) of the 
sneutrino field i). being zero; ii). being non-zero. In the paper
we would like to explore the more complex scenario with non-zero VEVs
for the sneutrino field of the third generation. 
In the sneutrino, as results, mixings of lepton-gaugino-Higgsino 
and slepton-Higgs etc, 
and a number of interesting phenomena are issued\cite{s13,s14}.
If the R-parity violation is originated from the vacuum only
without the breaking in the Lagrangian, then certain continual 
quantum number such as lepton number or else
must be associated to be broken,
so there will be certain physical Goldstone
particle occurring and a lot of 
phenomenological difficulties cannot be avoided
hence in the paper we will not discuss the case. 
Indeed one will see that in the present sneutrino, 
there is no physical Goldstone boson associating the breaking of R-parity. 
Here in the paper, taking an interesting example, we will consider a 
consequences of the bilinear slepton-Higgs R-parity violation 
terms on the Higgs masses and the mixed production $e^{+} e^{-} 
\rightarrow \tau^{\mp}
\tilde{\kappa}^{\pm}_{1}$ which is forbidden in the MSSM, and $\tilde{\kappa}
^{\pm}_{1}$ is to denote the lightest charginos.

The paper is organized as follows. Basic ingredients of the R-parity
violation MSSM with the explicit R-parity violation are briefly described
in Section II. In Section III, we will discuss the masses of 
CP-odd Higgs and charged Higgs sectors etc. The required massless Goldstone 
boson `eaten' by the electroweak gauge fields in unitary gauge is obtained
naturally, and the gauge-fixing terms in $'t$ Hooft-Feynman gauge
are derived. Furthermore, we take into account the effect of $e^{+} e^{-}
\rightarrow \tau^{\mp} \tilde{\kappa}^{\pm}_{1} $ in $e^{+} e^{-}$ colliders. In
Section V, we will present the numerical results calculated
under certain assumptions and discussions. In addition, we close
our discussions with short comments on certain implications of the
model for the other experiments.

\section{Minimal SUSY Model with Bilinear R-parity violation}

The supersymmetric Lagrangian is specified by the superpotential ${\cal W}$
that is given by\cite{s3},\cite{s15}:
\begin{eqnarray}
{\cal W} & = & \mu \varepsilon_{ij} \hat{H}_{i}^{1} \hat{H}_{j}^{2} +
\varepsilon_{ij} l_{IJ} \hat{H}_{i}^{1} \hat{L}_{j}^{I} \hat{R}^{J} +
\varepsilon_{ij} d_{IJ} \hat{H}_{i}^{1} \hat{Q}_{j}^{I} \hat{D}^{J}   \nonumber 
\\
  &  & +\varepsilon_{ij} u_{IJ} \hat{H}_{i}^{2} \hat{Q}_{j}^{I} \hat{U}^{J} +
\epsilon'^{I} \varepsilon_{ij}\hat{H}_{i}^{2} \hat{L}^{I}_{j}
\label{eq-1}
\end{eqnarray}
where $I, J = 1, 2, 3$ are generation indices, $i, j = 1, 2$ are SU(2)
indices, and $ \varepsilon $ is a completely antisymmetric $ 2\times 2$
matrix, with $ \varepsilon_{12} = 1$. The capital letters covered by a
symbol "hat" denote superfields: $\hat{Q}^{I}, \hat{L}^{I}, \hat{H}^{1}$, and $
\hat{H}^{2} $ being the SU(2) doublets with hyper-charges $ \frac{1}{3}, -1,
-1,$ and 1 respectively; $\hat{U}, \hat{D}$, and $\hat{R} $ being SU(2)
singlets with hyper-charges $-\frac{4}{3}, \frac{2}{3} $, and 2 respectively.
The couplings $u_{IJ}, d_{IJ}$, and $l_{IJ}$ are $3 \times 3$ Yukawa
matrices, and $\mu$, $\epsilon'^{I}$ are parameters with units of mass.
The first four terms in the superpotential are those as the MSSM, and
the last one is the R-parity violating term.

As MSSM, general and possible soft SUSY-breaking terms 
to break SUSY need to be introduced:
\begin{eqnarray}
{\cal L}_{soft} &=& -m_{H^{1}}^{2} H_{i}^{1*} H_{i}^{1}-m_{H^{2}}^{2} H_{i}^{2*}
H_{i}^{2}-m_{L^{I}}^{2} \tilde{L}_{i}^{I*} \tilde{L}_{i}^{I}  \nonumber  \\
  & & -m_{R^{I}}^{2} \tilde{R}^{I*} \tilde{R}^{I}-m_{Q^{I}}^{2} \tilde{Q}_{i}^{I*}
\tilde{Q}_{i}^{I}-m_{D^{I}}^{2} \tilde{D}^{I*} \tilde{D}^{I}  \nonumber  \\
 & &-m_{U^{I}}^{2} \tilde{U}^{I*} \tilde{U}^{I} + (m_{1} \lambda_{B} \lambda_{B}
 + m_{2} \lambda_{A}^{i} \lambda_{A}^{i}  \nonumber  \\
 & & + m_{3} \lambda_{G}^{a} \lambda_{G}^{a} + h.c.) + (B \mu \varepsilon_
{ij} H_{i}^{1} H_{j}^{2} + B_{1} \epsilon'^{I} \varepsilon_{ij} H_{i}^{2}
\tilde{L}_{j}^{I}  \nonumber  \\
 & & + \varepsilon_{ij} l_{sI} \mu H_{i}^{1} \tilde{L}_{j}^{I} \tilde{R}^{I}
+\varepsilon_{ij} d_{sI} \mu H_{i}^{1} \tilde{Q}_{j}^{I} \tilde{D}^{I}  
 \nonumber  \\
 & & + \varepsilon_{ij} u_{sI} \mu H_{i}^{2} \tilde{Q}_{j}^{I} \tilde{U}^{I}
 + h.c.)
 \label{eq-2}
\end{eqnarray}
where $m_{H^{1}}^{2}, m_{H^{2}}^{2}, m_{L^{I}}^{2}, m_{R^{I}}^{2}, m_{Q^{I}}^{2}, 
m_{D^{I}}^{2},$ and $m_{U^{I}}^{2}$ are the parameters with units of mass squared 
while $m_{1}, m_{2}, m_{3}$ denote the masses of the $SU(3)\times SU(2) \times U(1)$ gauginos
$\lambda_{G}^{a}, \lambda_{A}^{i}$ and $\lambda_{B}$, B and $B_{1}$ are free parameters with 
units of mass.

In order to eliminate unnecessary degrees of freedom, we assume that the
soft-breaking parameters and $\mu,\epsilon'^{I}(I=1,2,3)$ are real and
perform an operation that is the same as in the standard model by the
redefinition of the fields\cite{s16}:
\begin{eqnarray}
 \hat{Q}_{i}^{I} & \rightarrow & V_{Q_{i}}^{IJ} \hat{Q}_{i}^{J},  \nonumber \\
 \hat{U}^{I}  &  \rightarrow & V_{U}^{IJ} \hat{U}^{J} , \nonumber \\
 \hat{D}^{I}  &  \rightarrow & V_{D}^{IJ} \hat{D}^{J} , \nonumber \\
 \hat{L}_{i}^{I} & \rightarrow & V_{L_{i}}^{IJ} \hat{L}_{i}^{J},  \nonumber \\
 \hat{R}^{I}  &  \rightarrow & V_{R}^{IJ} \hat{R}^{J}
 \label{eq-3}
\end{eqnarray}
One can diagonalize the matrices $l_{IJ}, u_{IJ},$ and $ d_{IJ} $, the
superpotential has the form:
\begin{eqnarray}
{\cal W} & = & \mu \varepsilon_{ij} \hat{H}_{i}^{1} \hat{H}_{j}^{2} + l_{I}
\varepsilon_{ij} \hat{H}_{i}^{1} \hat{L}_{j}^{I} \hat{R}^{I} - 
u_{I}(\hat{H}^{2}_{1}
C^{JI*} \hat{Q}^{J}_{2}  \nonumber  \\
 & & - \hat{H}_{2}^{2} \hat{Q}^{I}_{1})\hat{U}^{I} - d_{I}(\hat{H}_{1}^{1} 
\hat{Q}
_{2}^{I} - \hat{H}_{2}^{1} C^{IJ}\hat{Q}_{1}^{J})\hat{D}^{I}  \nonumber  \\
 & & + \epsilon^{I} \varepsilon_{ij} \hat{H}_{i}^{2} \hat{L}_{j}^{I}
 \label{eq-4}
\end{eqnarray}
and the Kobayashi-Maskawa matrix C and $\epsilon^{I}$ have the definition as:
\begin{eqnarray}
  C & = & V_{Q_{2}}^{\dag} V_{Q_{1}}  \nonumber \\
  \epsilon^{I} & = & \epsilon'^{J} V_{L}^{JI}
  \label{eq-5}
\end{eqnarray}
and correspondingly the soft SUSY breaking sector has the form:
\begin{eqnarray}
{\cal L}_{soft} & = & -m_{H^{1}}^{2}H_{i}^{1*}H_{i}^{1} - m_{H^{2}}^{2} H_{i}^{2*}
H_{i}^{2}-m_{L^{I}}^{2} \tilde{L}_{i}^{I*} \tilde{L}_{i}^{I} - m_{R^{I}}^{2}\tilde{R}^{I*} 
\tilde{R}^{I}  \nonumber  \\
 &  & -m_{Q^{I}}^{2} \tilde{Q}_{i}^{I*} \tilde{Q}_{i}^{I} - m_{D^{I}}^{2}
\tilde{D}
 ^{I*} \tilde{D}^{I} - m_{U^{I}}^{2}\tilde{U}^{I*} \tilde{U}^{I} + (m_{1}
\lambda_{B}
\lambda_{B}  \nonumber  \\
 &  & + m_{2}\lambda_{A}^{i}\lambda_{A}^{i} + m_{3} \lambda_{G}^{a}\lambda_{
 G}^{a} + h.c.) + \{ B\mu \varepsilon_{ij}H_{i}^{1}H_{j}^{2} + B_{1}\epsilon^{
I}\varepsilon_{ij}H_{i}^{2}\tilde{L}_{j}^{I}  \nonumber \\
 & & + \varepsilon_{ij} l_{sI}\mu H_{i}^{1}\tilde{L}_{j}^{I}\tilde{R}^{I} +
  d_{sI}\mu (-H_{1}^{1}\tilde{Q}_{2}^{I} + C^{IK}H_{2}^{1}\tilde{Q}_{1}^{K})
 \tilde{D}^{I}  \nonumber  \\
 & &+ u_{sI}\mu (-C^{KI*}H_{1}^{2}\tilde{Q}_{2}^{I} + H_{2}^{2}\tilde{Q}_{1}
 ^{I})
 \tilde{U}^{I} + h.c.\}
 \label{eq-6}
 \end{eqnarray}
As pointed out at the above, 
from now on we take $\epsilon_{1} = \epsilon_{2} = 0$ always. In
this way, only $\tau$-lepton number is violated. 
The electroweak symmetry may be broken spontaneously in a general way
that the two Higgs doublets $H_{1}$, $H_{2}$, and the $\tau$-
sneutrino as well acquire vacuum expectation values(VEVs):
\begin{equation}
H^{1}=
\left(
\begin{array}{c}
\frac{1}{\sqrt{2}}(\chi_{1}^{0} + \upsilon_{1} + i\varphi_{1}^{0})  \\
H_{2}^{1}
\end{array}
\right)
\label{eq-7}
\end{equation}
\begin{equation}
H^{2}=
\left(
\begin{array}{c}
H^{2}_{1}  \\
\frac{1}{\sqrt{2}}(\chi_{2}^{0} + \upsilon_{2} + i\varphi_{2}^{0})
\end{array}
\right)
\label{eq-8}
\end{equation}
\begin{equation}
\tilde{L}_{3}=
\left(
\begin{array}{c}
\frac{1}{\sqrt{2}}(\chi_{3}^{0} + \upsilon_{3} + i\varphi_{3}^{0})  \\
\tilde{\tau}^{-}
\end{array}
\right)
\label{eq-9}
\end{equation}
It is easy to recognize the fact 
that the gauge bosons W and Z acquire masses given by $m_{W}^{2}=
\frac{1}{4}g^{2}\upsilon^{2}$ and $m_{Z}^{2}=\frac{1}{4}(g^{2}+g'^{2})
\upsilon^{2}$, where $ \upsilon^{2}=\upsilon_{1}^{2}+\upsilon_{2}^{2}
+\upsilon_{3}^{2}$ and $g, g'$ are coupling constants of SU(2) and U(1),
if one writes the rest sectors for the model relating to the gauge
fields. Let us introduce the following notation in 
spherical coordinates\cite{s3}:
\begin{eqnarray}
\upsilon_{1} & = & \upsilon \sin\theta_{\upsilon}\cos{\beta} \nonumber  \\
\upsilon_{2} & = & \upsilon \sin\theta_{\upsilon}\sin{\beta} \nonumber  \\
\upsilon_{3} & = & \upsilon \cos\theta_{\upsilon}
\label{eq-10}
\end{eqnarray}
which preserves the MSSM definition $\tan\beta=\frac{\upsilon_{2}}
{\upsilon_{1}}$. If furthermore 
the angle $\theta_{\upsilon}$ equals to $\frac{\pi}{2}$, this sector
will change back to the MSSM limit exactly. Note that in the literature
many authors choose a special direction $\theta_{\upsilon}=\frac{\pi}{2}$
by field redefinitions\cite{s21},
whereas we are considering the model with leptonic number either conserved
or violated in the soft SUSY breaking sector and only a bilinear R-parity
breaking term in superpotential thus here we leave the angle 
$\theta_{\upsilon}$ as a parameter to be determined phenomenologically.

The full scalar potential may be written as:
\begin{eqnarray}
V_{tree} & = & \sum_{i}|\frac{\partial W}{\partial A_{i}}|^{2} + V_{D}
+ V_{soft}  \nonumber  \\
  & = & V_{F} + V_{D} + V_{soft}.
\label{eq-11}
\end{eqnarray}
where $A_{i}$ denotes any one of the scalar fields in the theory, $V_{D}$
are the usual D-terms, $V_{soft}$ are the SUSY soft breaking terms give in
Eq.\ (\ref{eq-6}). Here, we do not consider 
the radiative corrections to the scalar potential at all.

The scalar term potential contains linear terms:
\begin{equation}
V_{linear} = t_{1}^{0}\chi_{1}^{0} + t_{2}^{0}\chi_{2}^{0} + t_{3}^{0}
\chi_{3}^{0}
\label{eq-12}
\end{equation}
where
\begin{eqnarray}
t_{1}^{0} & = & \frac{1}{4}(g^{2} + g'^{2})\upsilon_{1}(\upsilon_{1}^{2}
- \upsilon_{2}^{2} + \upsilon_{3}^{2}) +\frac{1}{2}|\mu|^{2}\upsilon_{1}
+\frac{1}{2}m_{H^{1}}^{2}\upsilon_{1} + \frac{1}{2}B\mu\upsilon_{2} +
\frac{1}{2}\epsilon_{3}\mu\upsilon_{3}  \nonumber  \\
t_{2}^{0} & = & -\frac{1}{4}(g^{2} + g'^{2})\upsilon_{2}(\upsilon_{1}^{2}
- \upsilon_{2}^{2} + \upsilon_{3}^{2})+\frac{1}{2}|\mu|^{2}\upsilon_{2}
 + \frac{1}{2}B\mu\upsilon_{1}+\frac{1}{2}m_{H^{2}}^{2}\upsilon_{2}
-\frac{1}{2}B_{1}\epsilon_{3}\upsilon_{3} + \frac{1}{2}\epsilon_{3}^{2}
\upsilon_{2}   \nonumber  \\
t_{3}^{0} & = & \frac{1}{4}(g^{2} + g'^{2})\upsilon_{3}(\upsilon_{1}^{2}
- \upsilon_{2}^{2} + \upsilon_{3}^{2}) + \frac{1}{2}m_{L^{3}}^{2}\upsilon_{3}
+\frac{1}{2}\epsilon_{3}^{2}\upsilon_{3} + \frac{1}{2}\epsilon_{3}\mu
\upsilon_{1} - \frac{1}{2}B_{1}\epsilon_{3}\upsilon_{2} .
\label{eq-13}
\end{eqnarray}
These $t_{i}^{0}, i={1, 2, 3}$ are the tree level tadpoles, and the VEVs of
the neutral scalar fields satisfy the condition $t_{i}^{0}=0, i={1, 2, 3}$ , 
we can obtain:
\begin{eqnarray}
m_{H^{1}}^{2} & = &-(|\mu|^{2} + \epsilon_{3}\mu\frac{\upsilon_{3}}{\upsilon
_{1}} + B\mu\frac{\upsilon_{2}}{\upsilon_{1}} + \frac{1}{2}(g^{2} + g'^{2})
(\upsilon_{1}^{2} - \upsilon_{2}^{2} + 
\upsilon_{3}^{2})) \nonumber  \\
m_{H^{2}}^{2} & =& -(|\mu|^{2} + \epsilon_{3}^{2} - B_{1}\epsilon_{3}\frac{
\upsilon_{3}}{\upsilon_{2}} + B\mu\frac{\upsilon_{1}}{\upsilon_{2}} -\frac{1
}{2}(g^{2} + g'^{2})(\upsilon_{1}^{2} - \upsilon_{2}^{2} + 
\upsilon_{3}^{2})) \nonumber  \\
m_{L_{3}}^{2} & = & -(\frac{1}{2}(g^{2} + g'^{2})(\upsilon_{1}^{
2} - \upsilon_{2}^{2} + \upsilon_{3}^{2}) + \epsilon_{3}^{2} + \frac{\epsilon
_{3}\mu\upsilon_{1}}{\upsilon_{3}} - B_{1}\epsilon_{3}\frac{\upsilon_{2}}{
\upsilon_{3}}) .
\label{eq-14}
\end{eqnarray}

An impact of the R-parity violation on the low energy phenomenology is twofold. 
Firstly, it leads the lepton number violation(LNV). Secondly, 
the bilinear R-parity violation term in the superpotential and that
in soft breaking terms generate the non-zero vacuum expectation value for the
sneutrino fields $ <\tilde{\nu}_{i}> \neq 0 $. As the consequences, not
only the neutrino-neutralino and 
electron-chargino mixing, but also various scalar mixings, 
such as those of the charged Higgs sector and the stau sector
etc are caused. In the discussions below, we always take the `new' 
parameters (those besides MSSM) at the weak 
interaction scale and impose on the restriction $m_{\nu_{\tau}} \leq 24 MeV$.

\section{Some Phenomenology of the BRpV Model}
\subsection{CP-odd neutral scalars and Charged Higgs-stau mixing}

The neutral scalar sector of the $\epsilon -$ model differs from that of the 
R-parity conserved MSSM: the Higgs bosons mix with the tau sneutrino. 
The CP-even sector is a
mixture of the real part of the $H_{1}^{1}$, $H_{2}^{2}$, and $\tilde{L_{1}^{3}}$, 
the mass matrix is given in Ref\cite{s3,s13}. Similarly, the CP-odd sector is 
a mixture of the imaginary part of the $H_{1}^{1}$, $H_{2}^{2}$, and 
$\tilde{L_{1}^{3}}$, after the mixing
there must be a linear combination corresponding to
the unphysical and massless Goldstone boson that is requested
for electroweak breaking. 

Let us see the fact precisely. In the original basis, 
where $\Phi_{odd}=(\varphi_{1}^{0}, \varphi_{2}^{0},\varphi_{3}^{0})$, the 
scalar potential contains the following mass term:
linear combination being the unphysical Goldstone boson. In the original basis
, where $\Phi_{odd}=(\varphi_{1}^{0}, \varphi_{2}^{0},\varphi_{3}^{0})$, the 
scalar potential contains the following mass term:
\begin{equation}
{\cal L}_{m}^{odd} = -\Phi_{odd}^{\dag}{\cal M}_{CP-odd}^{2}\Phi_{odd}
\label{eq-15}
\end{equation}
where the $3\times 3$ mass mixing matrix can be like this:
\begin{equation}
{\cal M}_{CP-odd}^{2} = 
\left(  \begin{array}{ccc}
r_{11} & -B\mu & \epsilon_{3}\mu  \\
-B\mu  & r_{22} & B_{1}\epsilon_{3} \\
\epsilon_{3}\mu & B_{1}\epsilon_{3} & r_{33}
\end{array}
\right)
\label{eq-16}
\end{equation}
with

\begin{eqnarray}
r_{11}=\frac{1}{2}(g^{2} + g'^{2})(\upsilon_{1}^{2}
-\upsilon_{2}^{2} + \upsilon_{3}^{2}) + |\mu|^{2} + m_{H^{1}}^{2} ,    \nonumber  \\
r_{22} = -\frac{1}{2}(g^{2} + g'^{2})(\upsilon_{1}^{2}-
\upsilon_{2}^{2} + \upsilon_{3}^{2}) + |\mu|^{2} + \epsilon_{3}^{2} + m_{H^{2}}^{2} , \nonumber \\
r_{33}=\frac{1}{2}(g^{2}+g'^{2})(\upsilon_{1}^{2}-\upsilon_{
2}^{2}+\upsilon_{3}^{2})+\epsilon_{3}^{2}+m_{L^{3}}^{2}. \nonumber  
\end{eqnarray}
 
Using Eq.\ (\ref{eq-14}), we can rewrite the matrix as below:
\begin{equation}
{\cal M}_{CP-odd}^{2} = 
\left(
\begin{array}{ccc}
-B\mu\frac{\upsilon_{2}}{\upsilon_{1}}-\epsilon_{3}\mu
\frac{\upsilon_{3}}{\upsilon_{1}} & -B\mu & \epsilon_{
3}\mu \\
-B\mu & B_{1}\epsilon_{3}\frac{\upsilon_{3}}{\upsilon_
{2}}-B\mu\frac{\upsilon_{1}}{\upsilon_{2}} & B_{1}
\epsilon_{3}  \\
\epsilon_{3}\mu &  B_{1}\epsilon_{3} & 
B_{1}\epsilon_{3}\frac{\upsilon_{2}}{\upsilon_{3}}-
\epsilon_{3} \mu \frac{\upsilon_{1}}{\upsilon_{3}}
\end{array}
\right)
\label{eq-17}
\end{equation}
The above matrix has an eigenstate:
\begin{eqnarray}
G^{0} &=& \sum_{i=1}^{3} Z^{odd}_{1,i}\varphi_{i}^{0}  \nonumber \\
      &=&\frac{1}{\upsilon}(\upsilon_{1}\varphi_{1}^{0} - \upsilon_{2}
\varphi_{2}^{0} + \varphi_{3}^{0})  \nonumber  \\
  &=&\sin\theta_{\upsilon} \cos\beta \varphi_{1}^{0} - \sin\theta_{\upsilon}
\sin\beta \varphi_{2}^{0} + \cos\theta_{\upsilon} \varphi_{3}^{0} .
\label{eq-18}
\end{eqnarray}
which is corresponding to 
the massless Goldstone boson which will disappear if the unitary gauge
is taken. 
The other two mass-eigenstates can be written as:
\begin{equation}
A_{i}^{0}(i=1,2) = \sum_{j=1}^{3}Z^{odd}_{i+1,j}\varphi_{j}^{0}
\label{eq-19}
\end{equation}
where $Z^{odd}_{i,j}(i, j=1, 2, 3)$ is the transformation matrix that rotates 
from the original basis into the mass-eigenstates. As we expected, 
all the $A_{i}^{0}(i=1,2)$ acquire masses.

In the model the complex scalar 
$H_{2}^{1*}$, $H_{1}^{2}$ mix with the left and right $\tau$-slepton. 
In the original basis, where $\Phi_{c}=(H_{2}^{1*}, H_{1}^{2}, 
\tilde{\tau}^{*}_{L}, \tilde{\tau}_{R})$, the scalar potential contains the 
following masses term:
\begin{equation}
{\cal L}_{m}^{C} = -\Phi_{c}^{\dag}{\cal M}_{c}^{2}\Phi_{c}
\label{eq-20}
\end{equation}
where the $ 4 \times 4$ mass matrix of the charged scalar sector can be divided
into three components for the model\cite{s13,s17} 
\begin{eqnarray}
{\cal M}^{2}_{c} = 
\left(
\begin{array}{cc}
M_{H}^{2} & 0  \\
0  & 0
\end{array} 
\right)
+\left(
\begin{array}{cc}
0 &  0  \\
0 & M_{\tilde{\tau}}^{2}
\end{array} 
\right)
+\left(
\begin{array}{cc}
0 & M_{\epsilon}^{2}  \\
M_{\epsilon}^{2\dag} & 0
\end{array} 
\right)  \nonumber
\end{eqnarray}
 where $M_{H}^{2} $ is
$2 \times 2$ charged Higgs masses matrix for MSSM and $M_{\tilde{\tau}}^{2}$ is 
the $2 \times 2$ stau masses matrix. The $M_{\epsilon}^{2} $ component 
does not present in MSSM, which cases a mixing
of the charged $H_{2}^{1*}$, $H_{1}^{2}$ and the $\tau$-slepton sector. 
The matrix ${\cal M}_{c}^{2}$ can be obtained in the model (here the matrix
is too big to be written precisely so we write it by each element individually):
\begin{eqnarray}
{\cal M}_{c 1,1}^{2} & = & g^{2}\upsilon_{1}^{2} - \frac{g^{2}-g'^{2}}{2}(
\upsilon_{1}^{2}-\upsilon_{2}^{2} + \upsilon_{3}^{2}) + |\mu|^{2} + \frac{
1}{2}\upsilon_{3}^{2}l_{(I=3)}^{2} + m_{H^{1}}^{2}  \nonumber  \\
   & = & g^{2}(\upsilon_{2}^{2}-\upsilon_{3}^{2}) + \frac{1}{2}\upsilon_{3}^{2}l_{(I=3)}^
{2} -\epsilon_{3}\mu\frac{\upsilon_{3}}{\upsilon_{1}}-B\mu\frac{\upsilon_{2}}{
\upsilon_{1}}, \nonumber  \\
{\cal M}_{c 1,2}^{2} &=&g^{2}\upsilon_{1}\upsilon_{2} - B\mu , \nonumber \\
{\cal M}_{c 1,3}^{2} &=& g^{2}\upsilon_{1}\upsilon_{3} +\epsilon_{3}\mu-\frac{
1}{2}l_{(I=3)}\upsilon_{1}\upsilon_{3}  , \nonumber \\
{\cal M}_{c 1,4}^{2} &=& \epsilon_{3}l_{(I=3)}\frac{\upsilon_{2}}{\sqrt{2}} -
l_{s(I=3)}\frac{\mu \upsilon_{3}}{\sqrt{2}} , \nonumber \\
{\cal M}_{c 2,2}^{2} &=& g^{2}\upsilon_{2}^{2} + \frac{1}{2}(g^{2}+g'^{2})
(\upsilon_{1}^{2}-\upsilon_{2}^{2}+\upsilon_{3}^{2}) + |\mu|^{2} + m_{H^{2}}^{
2}   \nonumber  \\
 & = & g^{2}(\upsilon_{1}^{2}+\upsilon_{3}^{2}) + B_{1}\epsilon_{3}
\frac{\upsilon_{3}}{\upsilon_{2}} - B\mu\frac{\upsilon_{1}}{\upsilon_{2}}, \nonumber \\
{\cal M}_{c 2,3}^{2} &=& g^{2}\upsilon_{2}\upsilon_{3} + B_{1}\epsilon_{3} ,
\nonumber  \\
{\cal M}_{c 2,4}^{2} &=& \frac{l_{(I=3)}}{\sqrt{2}}\mu\upsilon_{3} + \frac{
l_{(I=3)}}{\sqrt{2}}\epsilon_{3}\upsilon_{1} ,\nonumber \\
{\cal M}_{c 3,3}^{2} &=& g^{2}\upsilon_{3}^{2} + \frac{1}{2}(g^{2}+g'^{2})
(\upsilon_{1}^{2}-\upsilon_{2}^{2}+\upsilon_{3}^{2}) +\epsilon_{3}^{2}
\frac{l_{(I=3)}}{2}\upsilon_{1}^{2} + m_{L^{3}}^{2}  \nonumber \\
 & = & g^{2}(\upsilon_{2}^{2}-\upsilon_{1}^{2})-\epsilon_{3}\frac{\mu\upsilon_{1}}{
\upsilon_{3}} + B_{1}\frac{\epsilon_{3}\upsilon_{2}}{\upsilon_{3}} + 
\frac{l_{(I=3)}^{2}}{2}\upsilon_{1}^{3}, \nonumber \\
{\cal M}_{c 3,4}^{2} &=& \frac{1}{\sqrt{2}}l_{(I=3)}\mu\upsilon_{2}+\frac{1}{
\sqrt{2}}l_{s(I=3)}\mu\upsilon_{1} , \nonumber  \\
{\cal M}_{c 4,4}^{2} &=& -g'^{2}(\upsilon_{1}^{2}-\upsilon_{2}^{2}+\upsilon
_{3}^{2})+\frac{1}{2}l_{(I=3)}^{2}(\upsilon_{1}^{2} + \upsilon_{3}^{2})
+ m_{R^{3}}^{2}   \nonumber  \\
{\cal M}_{c 2,1}^{2} &=& {\cal M}_{c 1,2}^{2} ,\nonumber \\
{\cal M}_{c 3,1}^{2} &=& {\cal M}_{c 1,3}^{2} ,\nonumber \\
{\cal M}_{c 4,1}^{2} &=& {\cal M}_{c 1,4}^{2} ,\nonumber \\
{\cal M}_{c 3,2}^{2} &=& {\cal M}_{c 2,3}^{2} ,\nonumber \\
{\cal M}_{c 4,2}^{2} &=& {\cal M}_{c 2,4}^{2} ,\nonumber \\
{\cal M}_{c 4,3}^{2} &=& {\cal M}_{c 3,4}^{2} 
\label{eq-21}
\end{eqnarray}
where the Eq.\ (\ref{eq-14}) is used sometimes.
This matrix has an eigenstate:
\begin{eqnarray}
G^{+} &=& \sum_{i=1}^{4} Z_{1,i}^{c} \Phi^{i}_{c}  \nonumber  \\
      &=& \frac{1}{\upsilon}(\upsilon_{1} H_{2}^{1*} - \upsilon_{2} H_{1}^{2}
         +\upsilon_{3}\tilde{\tau}_{L}^{*})  \nonumber  \\
      &=& \sin\theta_{\upsilon}\cos\beta H_{2}^{1*}-\sin\theta_{\upsilon}\sin\beta
       H_{1}^{2}+\cos\theta_{\upsilon}\tilde{\tau}_{L}^{*}
\label{eq-23}
\end{eqnarray}
with zero eigenvalue, and being the massless charged Goldstone boson
it will be absorbed by W bosons and disappear in the physical
(unitary) gauge. The other three eigenstates $H^{+}, \tilde{\tau}_{1}, 
\tilde{\tau}_{2} $ can be expressed as:
\begin{eqnarray}
H^{+} &=& \sum_{i=1}^{4} Z_{2,i}^{c} \Phi^{i}_{c} ,\nonumber  \\
\tilde{\tau}_{1} &=& \sum_{i=1}^{4} Z_{3,i}^{c} \Phi^{i}_{c} ,\nonumber  \\
\tilde{\tau}_{2} &=& \sum_{i=1}^{4} Z_{4,i}^{c} \Phi^{i}_{c} .
\label{eq-24}
\end{eqnarray}

If a process is calculated only up to the tree approximation
in a spontaneously broken gauge theory, the most convenient
choice is to take the unitary gauge in which the unphysical Goldstone 
bosons will be absent in the Lagrangian and Feynman rules. 
Whereas when calculating
higher order corrections, it is convenient to choose a renormalizable
gauge, commonly the so-called $'$t Hooft-Feynman gauge
is favored\cite{s8}, in which the Goldstone fields 
appear explicitly. For
our later calculations, the appropriate choice for gauge fixing:
\begin{eqnarray}
{\cal L}_{GF} &=& -\frac{1}{2\xi}(\partial^{\mu}A_{\mu}^{3}+\xi m_{Z}\cos\theta_{W}
G^{0})^{2}-\frac{1}{2\xi}(\partial^{\mu}B_{\mu}-\xi m_{z}\sin\theta_{W} G^{0})
^{2}-  \nonumber  \\
 & & \frac{1}{2\xi}(\partial^{\mu}A_{\mu}^{1}+\frac{i}{\sqrt{2}}\xi m_{W}(
G^{+}-G^{-}))^{2}-  \nonumber  \\
 & & \frac{1}{2\xi}(\partial^{\mu}A_{\mu}^{2}-\frac{1}{\sqrt{2}}\xi m_{W}(
G^{+}+G^{-}))^{2}  \nonumber \\
 &=& \{-\frac{1}{2\xi}(\partial^{\mu}Z_{\mu})^{2}-\frac{1}{2\xi}(\partial^{
\mu}F_{\mu})^{2}-\frac{1}{\xi}(\partial^{\mu}W_{\mu}^{+})(\partial^{\mu}W_{
\mu}^{-})\} - \nonumber \\
 & & \{m_{z}G^{0}\partial^{\mu}Z_{\mu} + im_{W}(G^{+}\partial^{\mu}W_{\mu}^{-}
-G^{-}\partial^{\mu}W_{\mu}^{+}) \} - \nonumber \\
 & & \{ \frac{1}{2}\xi m_{Z}^{2}(G^{0})^{2} - \xi m_{W}^{2}G^{+}G^{-} \}
 \label{eq-25}
\end{eqnarray}
is taken. Here
\[ 
\cos^{2}\theta_{W} = \frac{m_{W}^{2}}{m_{Z}^{2}} 
\]
and with $G^{0}, G^{\pm}$ are defined as above. the first part of the above 
expression
is identical to the usual gauge-fixed terms; the second part cancels 
the off-diagonal vertices for Higgs-bosons-gauge-boson remaining 
in the Lagrangian after
symmetry breaking; and the third part `gives' masses to the Goldstone bosons
in the gauge.

\subsection{The Mixed Production $e^{+}e^{-} \rightarrow \tilde\kappa_{1}^{\pm}
\tau^{\mp}$ in the $e^{+}e^{-}$ Colliders} 

\begin{figure}
\setlength{\unitlength}{1mm}
\begin{picture}(160,200)
\put(-30,-20){\includegraphics{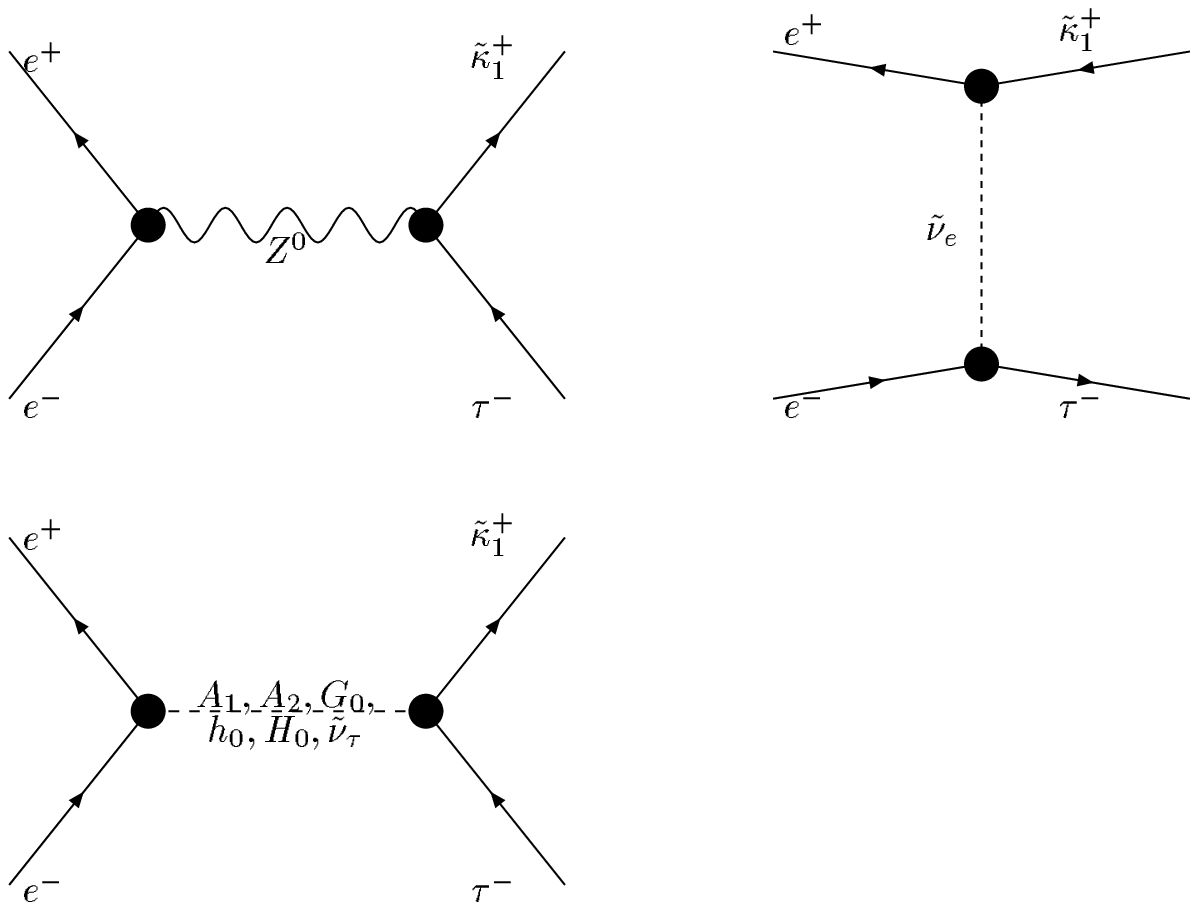}}
\end{picture}
\caption{The Feynman-diagrams for mixing production $e^{-}e^{+} \rightarrow
\tilde{\kappa}_{1}^{+} \tau^{-}$}
.\label{fig1}
\end{figure}

Similarly to the Higgs bosons, charginos mix with the $\tau$ lepton and
form a set of the charged fermions 
$\tau^{-}, \tilde{\kappa}_{1}^{-}, \tilde{\kappa}_{2}^{-}$ \cite{s13,s19}.
In the original basis where $\psi^{+T}=(-i\lambda^{+}, \tilde{H}_{2}^{1}, \tau_{
R}^{+})$ and $\psi^{-T}=(-i\lambda^{-}, \tilde{H}_{1}^{2}, \tau_{L}^{-})$, the 
charged fermion mass terms in the Lagrangian are:
\begin{equation}
{\cal L}_{m}=-\psi^{-T}{\cal M}_{f} \psi^{+}
\label{eq-26}
\end{equation}
with the mass matrix given by\cite{s13,s19}:
\begin{equation}
{\cal M}_{f} = 
\left(
\begin{array}{ccc}
\displaystyle 2m_{2} & \frac{e\upsilon_{2}}{\sqrt{2}S_{W}} & 0  \\
\displaystyle 
\frac{e\upsilon_{1}}{\sqrt{2}S_{W}} & \mu & \frac{l_{(I=3)}\upsilon_{3}}{\sqrt
{2}}  \\
\displaystyle 
\frac{e\upsilon_{3}}{\sqrt{2}S_{W}} & \epsilon_{3} & \frac{l_{(I=3)}\upsilon_{1}
}{\sqrt{2}}  
\end{array}
\right)
\label{eq-27}
\end{equation}
where $S_{W} = \sin\theta_{W}$ and $\lambda^{\pm}=\frac{\lambda_{A}^{1}\mp i\lambda_{A}^{2}}{
\sqrt{2}}$. 
Thus two mixing matrices $Z^{+}$ and $ Z^{-} $ appear, and they
are defined by the condition that the product $ (Z^{+})^{T}{\cal M}_{f} Z^{-}$
should be a diagonal matrix:
\begin{equation}
(Z^{+})^{T}{\cal M}_{f} Z^{-} =
\left(
\begin{array}{ccc}
m_{\tau} & 0 & 0 \\
0 & m_{\tilde{\kappa}^{-}_{1}} & 0 \\
0 & 0 & m_{\tilde{\kappa}^{-}_{2}}
\end{array}
\right)
\label{eq-28}
\end{equation}
The unitary matrices $Z^{+}$ and $Z^{-}$ are not uniquely specified if changing 
their relative phases and the order of the eigenvalues. It is possible to 
choose $m_{\tau}, m_{\tilde{\kappa}_{i}}$ positive and 
to have the order $ m_{\tilde{\kappa}_{2}^{-}}
\geq m_{\tilde{\kappa}_{1}^{-}} \geq m_{\tau} $, and we do so only for
fixing the irrelevant freedoms. Due to the 
mixing between $\tau$ and charginos, it is possible to occur the mixed 
production $ e^{+}e^{-} 
\rightarrow \tilde{\kappa}_{1}^{\pm} \tau^{\mp} $ which is forbidden in
MSSM. In the present
model, the Feynman diagrams that contribute to the lowest-order amplitude are
given in Fig.\ \ref{fig1}, and the contribution is mainly from the 
s-channel with the exchange of Z-gauge bosons and t-channel with 
the exchange of sneutrinos, whereas the contribution 
due to exchange of the other scalars is small, except crossing 
their corresponding resonances respectively.

\section{Numerical Results and Discussions}

Throughout the paper, we consider all the independent parameters to
take the values at the weak interaction scale. As known, there are 
many parameters in the model needed to be fixed still, so for
simplicity but not losing the key features, in the numerical evaluation
we assume there are some constraints among the parameters defined 
in Eqs.\ (\ref{eq-4}, \ref{eq-5}) as follow(all are at the weak 
interaction scale):
\begin{eqnarray}
\frac{l_{(I=3)}}{l_{s(I=3)}} = \frac{d_{(I=3)}}{d_{s(I=3)}} =
\frac{u_{(I=3)}}{u_{s(I=3)}},  \nonumber  \\
B  = -B_{1} = \frac{l_{s(I=3)}\mu}{l_{(I=3)}}-1 ,  \nonumber  \\
m_{H^{1}}^{2} = m_{H^{2}}^{2} = m_{L^{3}}^{2} = 
m_{R^{3}}^{2} = m_{Q^{3}}^{2} = m_{U^{3}}^{2} =m_{D^{3}}^{2},  \nonumber \\
m_{3} = m_{2} = m_{1} = M_{\frac{1}{2}}
\label{eq-29}
\end{eqnarray} 
With these constraints and Eq.\ (\ref{eq-14}) together, only four free 
parameters in this model are left. Hence we may choose $\tan\beta, \upsilon_{3},
\epsilon_{3}$ and $ M_{\frac{1}{2}}$ to be the four.

As for the other parameters that used in the numerical evaluation, we take
$\alpha \equiv \frac{e^{2}}{4\pi} = \frac{1}{128}$, $m_{e}=0.51MeV$, 
$m_{\tau}=1.78 GeV$,
$M_{Z}=91.19 GeV$, $M_{W}=80.23 GeV$. 
 
In order to find out the allowed region in the parameter space, one has to 
take all the experimental constraints into account. 
First, we would like to note that $\epsilon_{3},
M_{\frac{1}{2}}$, and $\upsilon_{3}$ are the three free parameters which enter
into the chargino and neutralino mass matrices. Then a very strong restriction 
on the parameters comes from the fact that the $\tau$ mass 
has been measured very precisely\cite{s20}, therefore, for any 
combination of the $\epsilon_{3}$, $M_{\frac
{1}{2}}$
and $\upsilon_{3}$, the lowest eigenvalue of Eq.\ (\ref{eq-27}) should agree with 
$m_{\tau}$.
Also, $\nu_{\tau}$ has a laboratory upper bound on its mass 
$m_{\nu_\tau}\leq 24MeV$. These
two restrictions, together with the positive-definite
condition for the Higgs mass matrices, the
restrict the allowed parameter space very seriously. Furthermore, since 
we are interested in relatively light charginos, thus in the numerical
calculation we take $M_{\frac{1}{2}} \sim 300GeV$ so small.

\begin{figure}
\setlength{\unitlength}{1mm}
\begin{picture}(160,200)
\put(-30,-20){\includegraphics{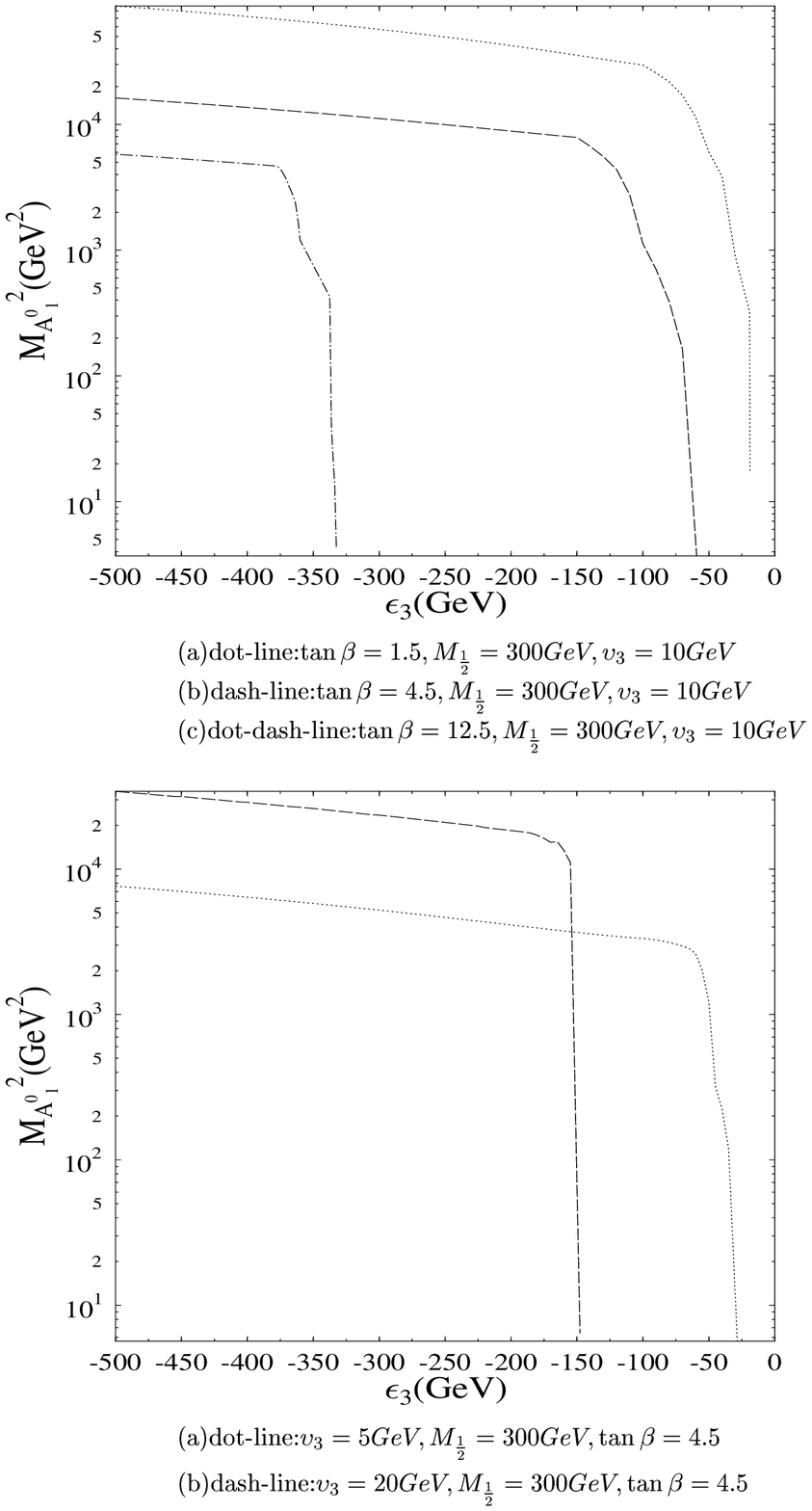}}
\end{picture}
\caption{the $M_{A_{1}^{0}}^{2}$ vary with $\epsilon_{3}$}
.\label{fig2}
\end{figure}

\begin{figure}
\setlength{\unitlength}{1mm}
\begin{picture}(160,200)
\put(-30,-20){\includegraphics{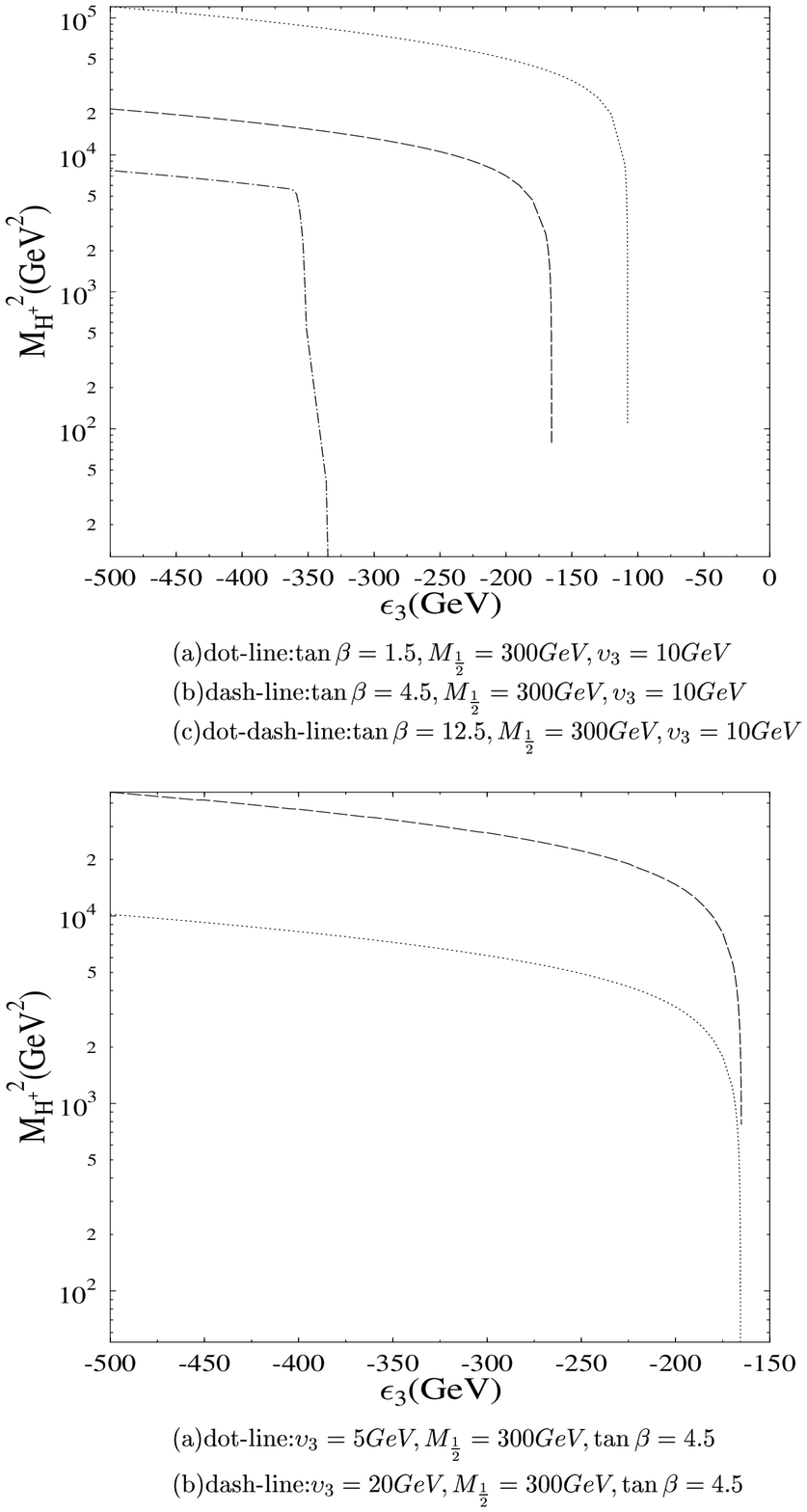}}
\end{picture}
\caption{the $M_{H^{+}}^{2}$ vary with $\epsilon_{3}$}
.\label{fig3}
\end{figure}

\begin{figure}
\setlength{\unitlength}{1mm}
\begin{picture}(160,220)
\put(-30,-20){\includegraphics{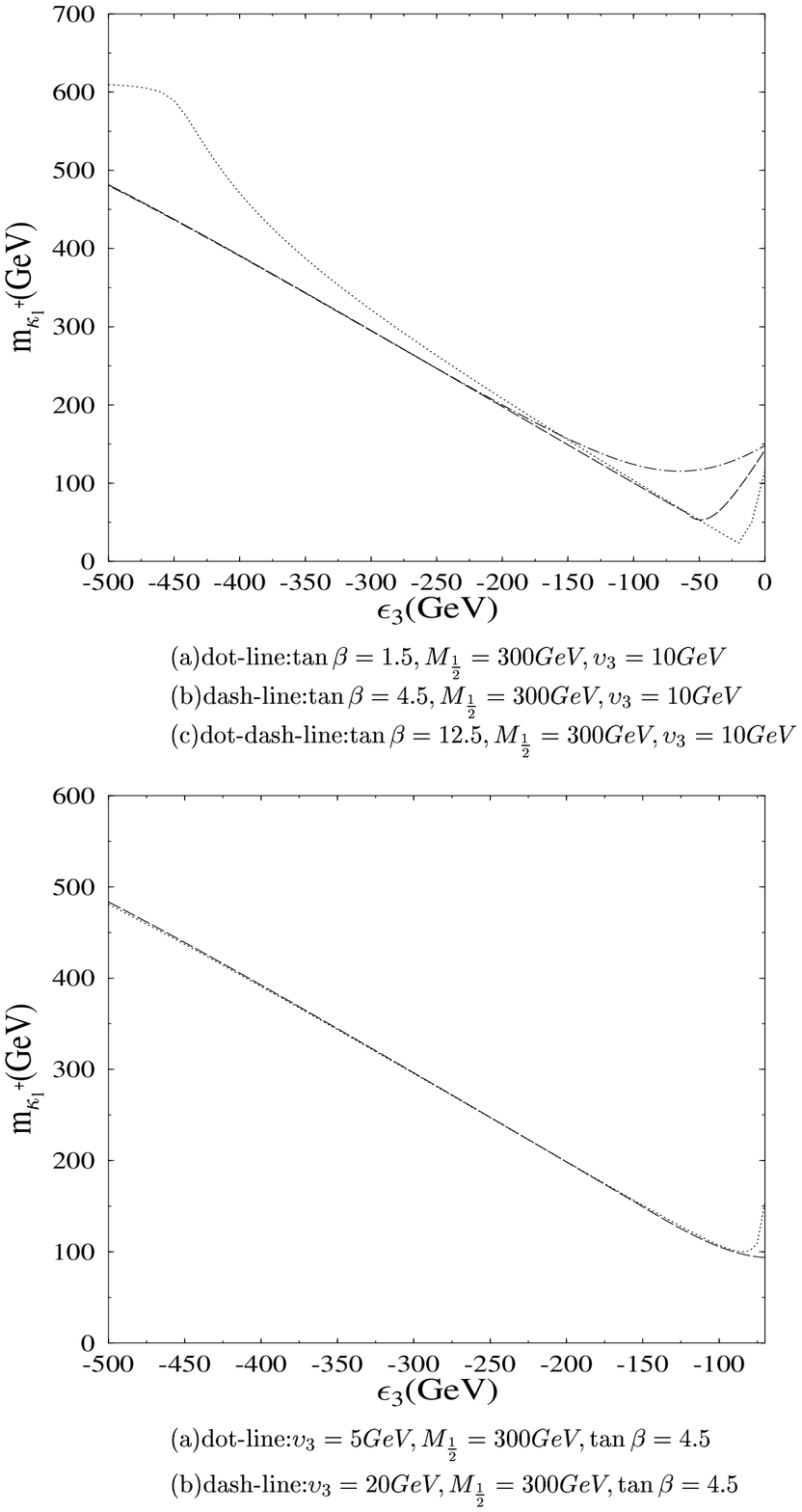}}
\end{picture}
\caption{the $M_{\tilde{\kappa}_{1}^{+}}$ varied with $\epsilon_{3}$}
.\label{fig4}
\end{figure}

Fig.\ \ref{fig2},Fig.\ \ref{fig3} show mass squared of the lightest CP-odd Higgs
 and mass squared of the
lightest charged Higgs varied with the parameter $\epsilon_{3}$. With the
assumption Eq.\ (\ref{eq-29}), the $\epsilon_{3}$ must be less than zero. The main 
point to
note is that $M_{H^{+}}^{2}$ can be lower than the expected value in MSSM
due to the fact that in the model the sneutrino acquires nonzero VEVs so
a negative contribution from the R-parity violating stau-Higgs mixing 
results in. It in fact is controlled by the parameter $\epsilon_{3}$ 
and $\upsilon_{3}$. 
From Fig.\ \ref{fig3}, one may see that the charged Higgs mass may turn
to small when $\epsilon_{3}$ approaches to a certain value, varying with 
the other parameters taken. It is also because we have 
made the assumption the Eq.\ (\ref{eq-29}). From the scalar 
potential Eq.\ (\ref{eq-11}),
we have:
\begin{eqnarray}
V_{tree} & = & m_{L^{3}}^{2}\tilde{L}_{1}^{3*} \tilde{L}_{1}^{3} + \epsilon_{3}^{2}
\tilde{L}^{3*}_{1}\tilde{L}^{3}_{1} + (B_{1}\epsilon_{3}H_{2}^{2}\tilde{L}_{1}^{3} 
+ h.c.) - (\mu\epsilon_{3}H_{1}^{1}\tilde{L}_{1}^{3} + h.c.) +  \nonumber  \\
 & & \frac{g_{1}^{2}+g_{2}^{2}}{8}\{ (\tilde{L}_{1}^{3*}\tilde{L}_{1}^{3})^{2} +
2\tilde{L}_{1}^{3*}\tilde{L}_{1}^{3}H_{1}^{1*}H_{1}^{1} - 2\tilde{L}_{1}^{3*}
\tilde{L}_{1}^{3}H_{2}^{2*}H_{2}^{2*} \} + \cdots
\end{eqnarray}
If the $\tau$-sneutrino has a non-zero vacuum expectation value, $\epsilon_{3}^{2}
+ m_{L^{3}}^{2}$ must be negative. Because we interest the case that $\epsilon_{3}$
parameter is real, so $m_{L^{3}}^{2}$ is a negative number. Under conditions
Eq.\ (\ref{eq-29}), $m_{R^{3}}^{2}$ is a negative number. Furthermore, from the
Eq.\ (\ref{eq-14}) and the relations in Eq.\ (\ref{eq-29}):
\[ B = -B_{1} = \frac{l_{s(I=3)}\mu}{l_{(I=3)}} -1 \]
 and \[ m_{H^{1}}^{2} = m_{H^{2}}^{2} = m_{L^{3}}^{2} \]
the mass matrix of charged Higgs depends on the $\epsilon_{3}$ 
in a very complicated manner. The two reasons make the mass matrix of 
charged Higgs is not positive-definite when $\epsilon_{3}$ approaches 
to the value when $\tan\beta$, $\upsilon_{3}$ and $M_{\frac{1}{2}}$ are given. 
We can
understand the Fig.\ \ref{fig2} in a similar way. Fig.\ \ref{fig4} shows the mass
of the lightest chargino varied with $\epsilon_{3}$, the minimum of $m_{
\tilde{\kappa}_{1}}$ is about $ 100 GeV$. If we don't consider the constraint 
Eq.\ (\ref{eq-29}), the value of $\epsilon_{3}$ can be larger than zero, this case has been 
discussed by Ref\cite{s13} and Ref\cite{s17}.

\begin{figure}
\setlength{\unitlength}{1mm}
\begin{picture}(160,160)
\put(-30,-10){\includegraphics{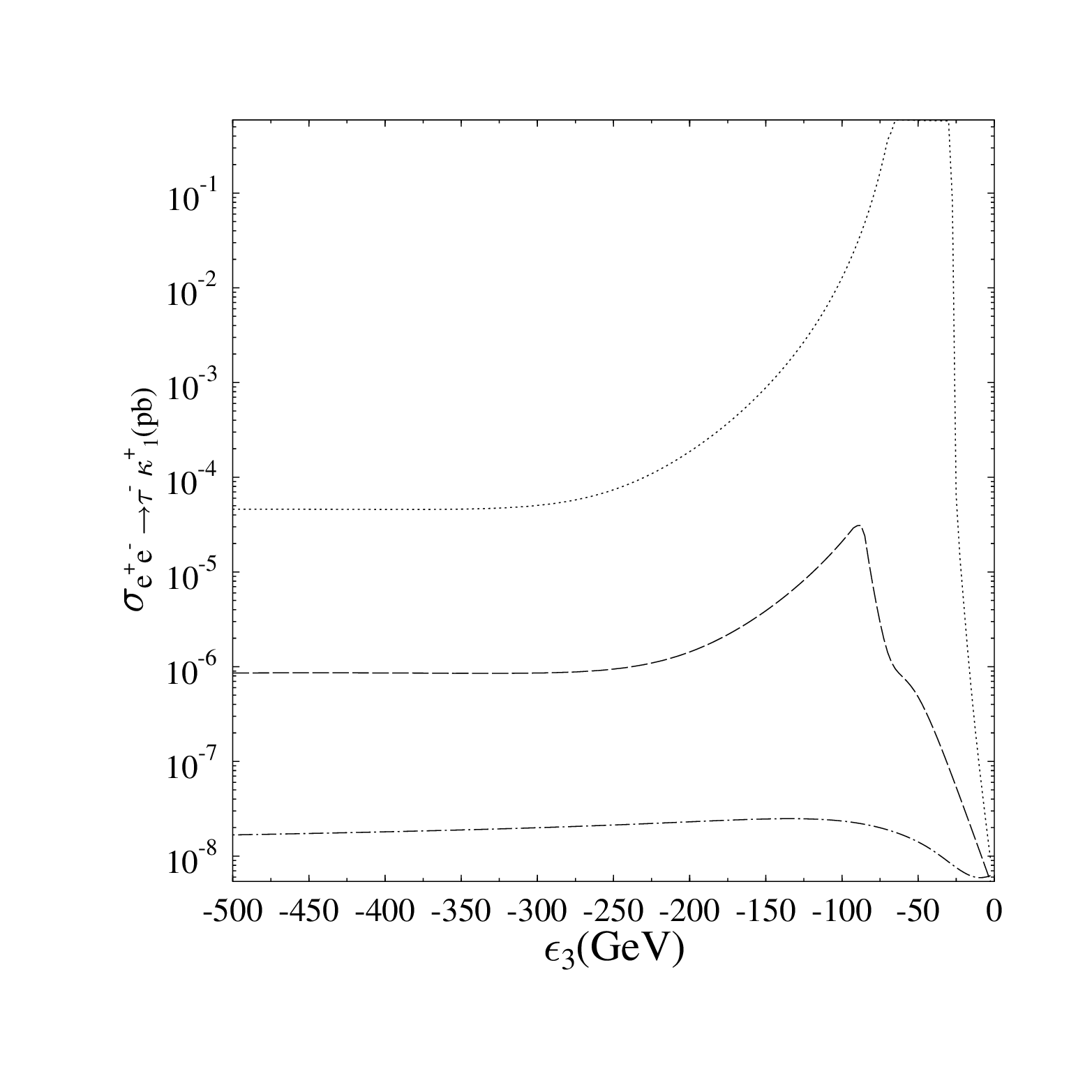}}
\end{picture}
\caption{The lowest-order cross sections of mixing production $\tau^{\mp}
\tilde{\kappa}_{1}^{\pm}$ against $\epsilon_{3}$ with \\
(a)dot-line:$\tan\beta=4.5, M_{\frac{1}{2}}=300 GeV, \upsilon_{3}=5 GeV, 
\sqrt{S}=1000 GeV$;\\ 
(b)dash-line:$\tan\beta=12.5, M_{\frac{1}{2}}=300 GeV, \upsilon_{3}=5 GeV, 
\sqrt{S}=1000 GeV$;\\
(c)dot-dash-line:$\tan\beta=40.5,M_{\frac{1}{2}}=300 GeV, \upsilon_{3}=5 GeV,
\sqrt{S}=1000 GeV.$ }
.\label{fig5}
\end{figure}

\begin{figure}
\setlength{\unitlength}{1mm}
\begin{picture}(160,160)
\put(-30,-10){\includegraphics{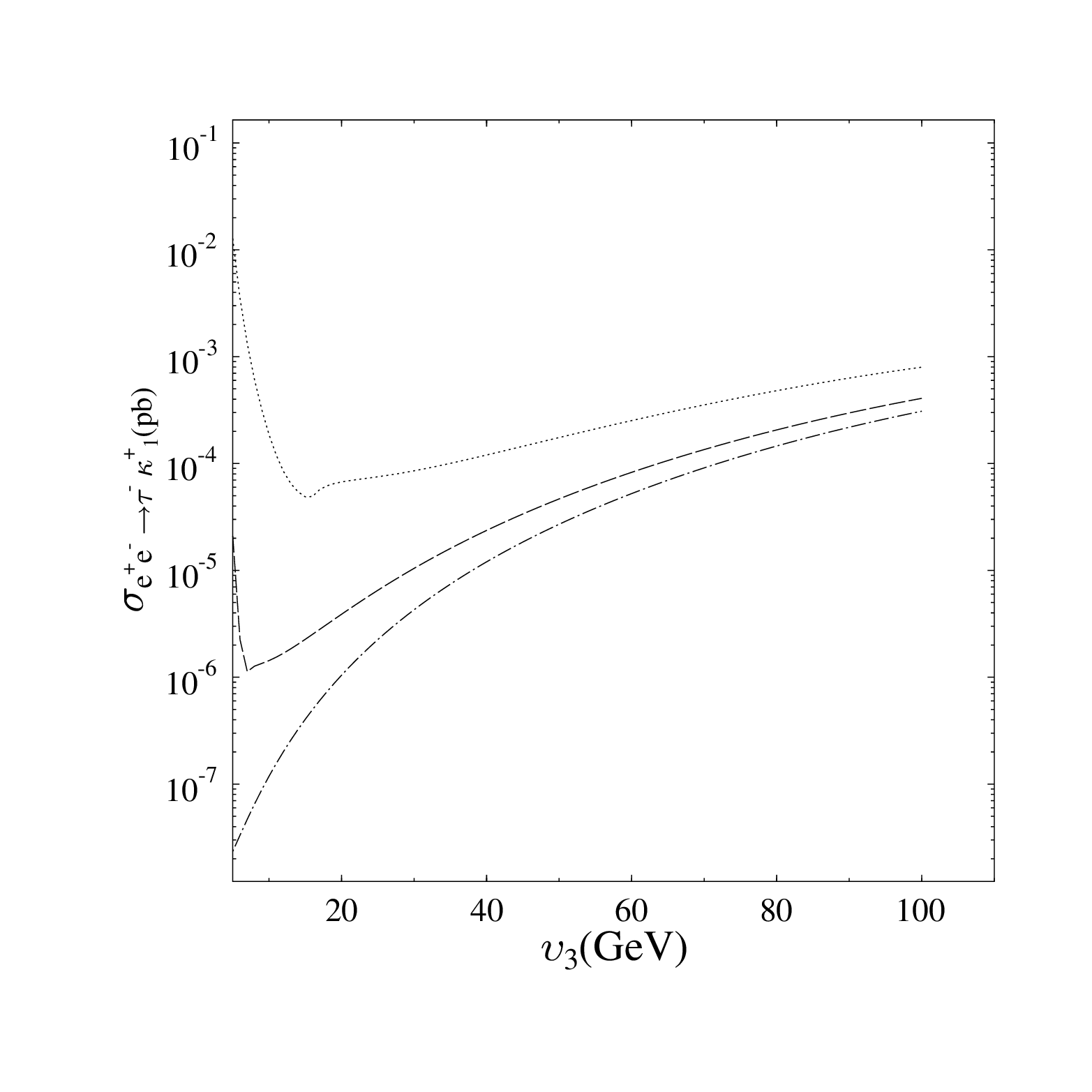}}
\end{picture}
\caption{The lowest-order cross sections of mixing production $\tau^{\mp}
\tilde{\kappa}_{1}^{\pm}$ against $\upsilon_{3}$ with \\
(a)dot-line:$\epsilon_{3}=-100 GeV, M_{\frac{1}{2}}=300 GeV, \tan\beta=4.5, 
\sqrt{S}=1000 GeV;$\\
(b)dash-line:$\epsilon_{3}=-100 GeV, M_{\frac{1}{2}}=300 GeV, \tan\beta=12.5,
\sqrt{S}=1000 GeV$;\\
(c)dash-dot-line:$\epsilon_{3}=-100 GeV, M_{\frac{1}{2}}=300 GeV, 
\tan\beta=40.5, \sqrt{S}=1000 GeV.$ }
.\label{fig6}
\end{figure}

Finally, let us discuss the mixing production $\tilde{\kappa}_{1}^{\pm} 
\tau^{\mp}$ 
as the typical consequences of the bilinear R-violating terms. In 
Fig.\ \ref{fig5},
we plot the $\sigma_{e^{-}e^{+} \rightarrow \tilde{\kappa}_{1}^{\pm} \tau^{\mp}}
$ against $\epsilon_{3}$(in GeV),
$\sigma_{e^{-}e^{+} \rightarrow \tilde{\kappa}_{1}^{\pm} \tau^{\mp}} \rightarrow
0$ when $|\epsilon_{3}| \rightarrow 0$. The $\sigma_{e^{-}e^{+} \rightarrow 
\tilde{\kappa}_{1}^{\pm} \tau^{\mp}} $ varied with $\upsilon_{3}$ is plotted in 
Fig.\ \ref{fig6}. From Fig. \ \ref{fig5} and Fig. \ \ref{fig6}, we find that
the $\sigma_{e^{-}e^{+} \rightarrow \tilde{\kappa}_{1}^{\pm} \tau^{\mp}}$
depend on the parameters $\tan\beta$ and $\upsilon_{3}$  strongly. If we 
release Eq.(29), the case is very involved.

In summary, it is shown that the Bilinear R-parity Violation
Model is one of the simplest extension of MSSM, in which the
R-parity violation is introduced by two folds: a violation term in
the Lagrangian and the VEVs of the sneutrino. In the model 
there are two massless Goldstone $G^{0}, G^{\pm}$, 
requested to be `eaten' by week bosons as the manner in SM 
and MSSM in the unitary physical gauge.
As a quite large value of $\epsilon$ and $\upsilon_{3}$ is allowed
in the model, so we are quite sure that with the parameters
one can find certain differences of the model from MSSM in phenomenology
at tree and/or one-loop level. Being a consequence
of the bilinear R-violation term, the 
$e^{-}e^{+} \rightarrow \tilde{\kappa}_{1}^{\pm}\tau^{\mp}$ can occur, 
and the cross section is typical in order $10^{-4}pb$ when 
$|\epsilon_{3}|$ is so large as $|\epsilon_{3}| \sim 100 GeV$. 

\vspace{2cm}
{\Large\bf Acknowledgement} This work was supported in part by the National Natural
Science Foundation of China and the Grant No. LWLZ-1298 of the Chinese Academy of
Sciences.

\begin{references}
\bibitem{s1}  H. Haber and G. Kane, Phys. Rep. 117, 75(1985); H. P. Nilles, Phys
. Rep. 110, 1(1984).
\bibitem{s2}  G. Farrar and P. Fayet, Phys. Lett. 76B, 575(1978): Phys. Lett.79B, 
442(1978).
\bibitem{s3}  M. A. Diaz, J. C. Romao and J. W. F. Valle, hep-ph/9706315.
\bibitem{s4}  L. Hall and M. Suzuki, Nucl. Phys. B 231, 419(1984).
\bibitem{s5}  T. Banks, T. Grossman, E. Nardi, Y. Nir, Phys. Rev. D 52, 5319(1995);
E. Nardi, Phys. Rev. D 55, 5772(1997).
\bibitem{s6}  F. de Campos, M. A. Garcia-Jareno, A. S. Joshipura, J. Rosiek, J. W.
F. Valle Nucl. Phys. B 451, 3(1995).
\bibitem{s7}  F. M. Borzumati, Y. Grossman, E. Nardi, and Y. Nir, WIS-96-21-PH,
hep-ph/9606251.
\bibitem{s8}  H.-P. Nilles and N. Polonsky, Nucl. Phys. B 484, 33(1997).
\bibitem{s9}  A. Faessler, S. Kovalenko, hep-ph/9712535.
\bibitem{s10}  C. Aulakh and R. Mohapatra, Phys.Lett. B119, 136(1983); G. G. Ross
and J. W. F. Valle, Phys. Lett. B151, 375(1985); J. Ellis, G. Gelmini, C.Jarlskog,
 G. G. Ross, and J. W. F. Valle, Phys. Lett. B150, 142(1985); A. Santamaria and
 J. W. F. Valle, Phys. Lett. B195, 423(1987); Phys. Rev. D39, 1780(1989); Phys. Rev.
Lett. 60, 397(1988); A. Masiero, J. W. F. Valle, Phys. Lett. B251, 273(1990).
\bibitem{s11}  J. C. Romao, A. Ioannissyan and J. W. F. Valle, Phys. Rev. D55, 427(
1997).
\bibitem{s12}  R. Hempfling, Nucl. Phys. B 478, 3(1996); B. de Carlos, P.L. White,
Phys. Rev.D 55, 4222(1997).
\bibitem{s13}  M. Nowakowski and A. Pilaftsis, Nucl. Phys. B461, 19(1996); A. 
Joshipra and M. Nowakowski, Phys. Rev. D51, 2421(1997).
\bibitem{s14}  S. Roy and B. Mukhopadhyaya, Phys. Rev. D55, 7020(1997).
\bibitem{s15}  M. A. Diaz, hep-ph/9711435.
\bibitem{s16}  J. Rosiek, Phys. Rev. D41, 3464(1990).
\bibitem{s17}  J. Ferrandis hep-ph/9802275.
\bibitem{s18}  C. Itzykson and J-B. Zuber, Quantum Field Theory(McGraw-Hill,New
York, 1980).
\bibitem{s19}  M. A. Diaz, hep-ph/9802407.
\bibitem{s20}  Particle Data Group. R. M. Barnett et al., Phys. Rev. D54, 1(1996).
\bibitem{s21}  S.Davidson, J. Ellis, Phys. Letts B390, 210(1997); S.Davidson, 
               J. Ellis, Phys. Rev. D56, 4182(1997).
\end{references}
\end{document}